\documentclass{article}  
\usepackage{lajolla2006}
\usepackage{epsfig}
\frompage{000} \topage{000}                                              
\def\rightmark{Radioactive beams at Argonne}
\def\leftmark{B. B. Back}
 
\title{The Radioactive Beam Program at Argonne} 
\authors{{B. B. Back}\\[2.812mm]
{\normalsize
Argonne National Laboratory\\ 
Argonne, Illinois 60439, USA\\
E-mail: Back@anl.gov\\[0.2ex] 
}}
 
\abstract{In this talk I will present selected topics of the ongoing radioactive beam program at Argonne and discuss the capabilities of the CARIBU radioactive ion production facility as well as plans for construction of a novel superconducting solenoid spectrometer.}

 
\begin{document}
 
\maketitle
\setcounter{page}{1}

\section{Introduction}\label{intro}

The availability and use of radioactive beams has recently become a central theme in nuclear- and astro-physics research. At present, there are several facilities devoted to nuclear physics studies using low-energy re-accelerated beams of radioactive species including the Cyclotron Research Centre at Louvain-la-Neuve, Belgium, Spiral at Ganil, Caen, France, TRIUMF in Vancouver, Canada, and the Holifield Radioactive Ion Beam Facility in Oak Ridge, USA among others. Currently, there are plans to enhance the capabilities for radioactive beams at several facilities, which will eventually allow for detailed studies of the structure of nuclei on the path of the astrophysical $r$-process and allow for fundamental nuclear structure studies of very neutron-rich nuclei. In anticipation of these future capabilities, a program of radioactive beam studies has been undertaken at the Argonne ATLAS accelerator over the last decade. In the following, some recent milestones of this program will be discussed as well as current upgrade plans, which will provide a substantial extension to present capabilities.

\section{Present program}

\subsection{Two-accelerator method}
Several techniques have been employed in order to obtain beams of radioactive nuclei. For relatively long-lived species it is possible to use normal ionization and acceleration techniques for material produced efficiently at another accelerator. This technique was employed for making $^{18}$F-beams used in the study of the $^{18}$F(p,$\alpha)^{15}$O reaction\cite{Rehm_18F}. In this case the 110 min half-life of $^{18}$F allowed for its production at a medical cyclotron at the University of Wisconsin in Madison, Wisconsin. After irradiation and chemical preparation, the material was quickly transported to Argonne by air and inserted in the ion source at the ATLAS accelerator. 

A study of the single particle structure near the doubly magic $^{56}$Ni nucleus\cite{Rehm_56Ni} was carried out with $^{56}$Ni beams using material produced via the $^{58}$Ni(p,p2n)$^{56}$Ni reaction with an intense 50 MeV proton beam from the Intense Pulsed Neutron Source (IPNS) at Argonne. In this work the single-neutron states in $^{57}$Ni were studied with the (d,p) reaction in inverse kinematics using the $^{56}$Ni beam.

Finally, a recent study of the $^{44}$Ti($^3$He,p) reaction~\cite{Macchi}, again studied in inverse kinematics, used source material produced at the Los Alamos Neutron Sciences Center (LANSCE) 

\subsection{In-flight production}

\begin{figure}[hbt]
\centerline{\epsfig{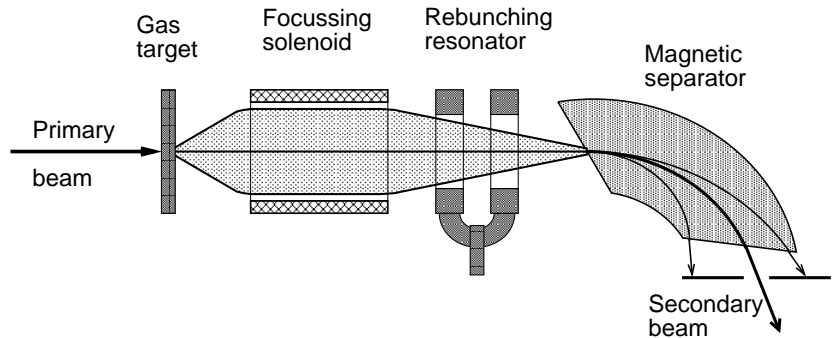}}
\caption{Schematic illustration of the in-flight radioactive beam setup at ATLAS. A solenoid provides focusing to the heavy reaction products, which are rebunched in energy by a resonator to minimize losses in the magnetic separator.}
\label{in-flight}
\end{figure}

In order to provide beams of shorter-lived isotopes, for which the above method is not feasible, an in-flight capability has been developed. This facility is based on the re-capture of heavy reaction products from inverse light-ion reactions, such as (d,p), (p,n) etc. using stable heavy-ion beams from the ATLAS accelerator and a target gas cell\cite{Harss}. A schematic illustration of this setup is shown in Fig.~\ref{in-flight}.

\begin{figure}[hbt]
\centerline{\epsfig{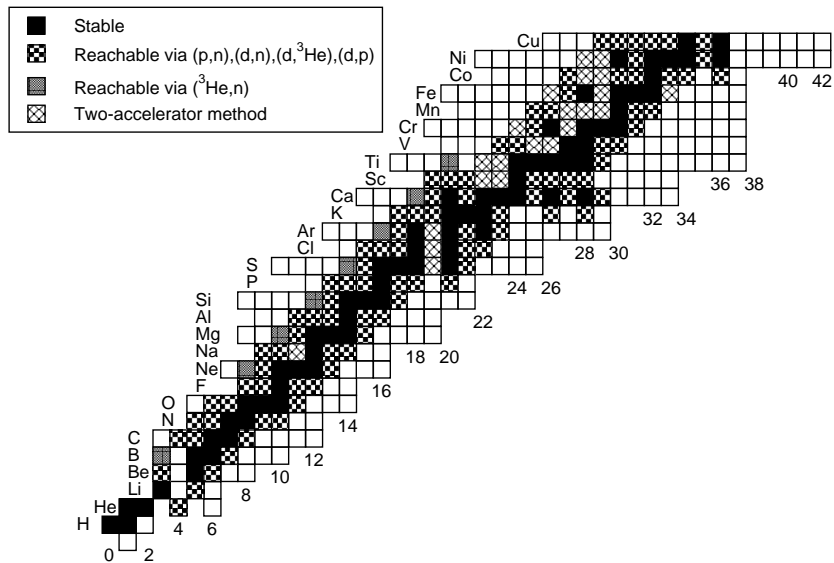}}
\caption{Radioactive beams that can be produces at ATLAS.}
\label{RadBeams}
\end{figure}

The radioactive beams obtained in this setup range in intensity up to $\sim10^6$ particles/sec, which are delivered to the experimental setup after transport through a high-transmission beam-line section. Beams produced in this fashion include $^6$He, $^8$Li,$^{8,12}$B,$^{10,11}$C,$^{14}$O,$^{16}$N,$^{17}$F,$^{19}$Ne,$^{20,21}$Na,$^{26}$Al,$^{37}$K~\cite{RadBeams}. Additional radioactive beams can be developed as required by the experimental program. Fig.~\ref{RadBeams} shows the range of beams that can be developed using light ion reactions 

\section{Recent experiments}

\subsection{The $^{12}C(\alpha,\gamma)$ reaction}

As one of the most important reaction in explosive stellar nucleosynthesis, the $^{12}$C($\alpha,\gamma$)$^{16}$O reaction is of great current interest. At issue is the contribution of a sub-threshold 1$^-$ state in $^{16}$O to the reaction rate, which is strongly dependent on its width. Several previous measurements~\cite{Mainz,Yale,Triumf} have studied this issue by population of excited 1$^-$ states in $^{16}$O by ground-state $\beta$ decay of $^{16}$N using Si detectors, which are sensitive to both $\beta$ particles and the subsequent $\alpha$'s emitted by the excited 1$^-$ states. To minimize the contribution of the $\beta$'s to the observed spectrum, this experiment employed a twin-ionization chamber viewing foils containing the implanted $^{16}$N beam produced in a $^2$H($^{15}$N,$^{16}$N)p reaction. This experimental method is still being refined but it is expected to lead to a much desired improvement in the accuracy of this important reaction rate~\cite{Tang}. 

\subsection{The $^6$He(d,p) reaction}

Recently, there has been substantial progress in {\it ab-initio} theoretical calculations of the properties of light nuclei (A$<$12) using the Greens Function Monte Carlo method~\cite{GFMC}. The results of this theoretical approach compare well with many of the known experimental properties of stable light nuclei, but it is clearly also of interest to confront them with data on unstable isotopes. This has been done in several cases with radioactive beams produced by the in-flight method at ATLAS. The two cases discussed here are both using $^6$He beams. 

In the first case, a $^6$He beam produced in-flight via the $^2$H($^7$Li,$^6$He)$^3$He primary reaction was used to study single neutron states in $^7$He via the $^2$H($^6$He,$^7$He)p reaction by detecting both the $^7$He decay products and the proton~\cite{Wuosmaa}. At issue is the existence of a broad $\Gamma$=750 keV 1/2$^-$ level at $E_x$=600 keV inferred by Meister {\it et al.}~\cite{Meister} from a $^8$He breakup reaction. The ATLAS experiment was carried out using an efficient Si detector setup including three large annular Si detectors to register protons emitted into backward angles and a large-area forward Si $\Delta$E-E telescope to identify the forward-going $\alpha$-particles from the subsequent $^7$He breakup channel. The data clearly excludes the existence of a 1/2$^-$ state of the reported width and excitation energy, a result that is in accord with both previous direct measurements and with theoretical expectations. 

\subsection{The charge radius of $^6$He}

Another important study carried out recently at ATLAS is the measurements of the nuclear charge radius of $^6$He atoms. The charge radius is sensitive to the detailed configuration of this system, which is expected to consist of two neutrons in a halo structure outside of the tightly bound $\alpha$-particle core. The charge radius is sensitive to the spatial correlation between the two halo neutrons. 

The measurement was carried out by precision laser spectroscopy on $^6$He atoms ($T_{1/2}$= 0.8 sec) which were produced in the reaction $^{12}$C($^7$Li,$^6$He)$^{13}$N using a hot graphite target and subsequently brought to rest in a magneto-optical trap~\cite{Wang}. A root-mean-square charge radius of $\langle r^2_c \rangle ^{1/2}= 2.054 \pm 0.056$ fm was measured. This represents a substantial improvement in comparison to previous results, which were derived from measurements of the total reaction cross section~\cite{Tanihata} or proton elastic scattering~\cite{Alkhazov}. The precision of the present work thus allows for critical tests of theories predicting the structure of light nuclei~\cite{GFMC}.

\section{CARIBU - a source of neutron-rich ions}

Fragments emitted in spontaneous fission of $^{252}$Cf have been the subject of numerous studies for many years. Because of the curvature of the line of $\beta$-stability, these fragments fall in the neutron-rich region for elements ranging from $^{83}$As to $^{166}$Tb with intensity maxima at $^{106}$Mo and $^{141}$Cs. Much information about neutron-rich nuclei produced in this fashion has been obtained by studying the subsequent $\gamma$-ray cascade in GAMMASPHERE and other modern $\gamma$-ray detectors. Much more nuclear structure information could be obtained, however, by using beams of these exotic nuclei in Coulomb excitation, inelastic scattering, few nucleon transfer, and fusion reactions. 

\begin{figure}[hbt]
\centerline{\epsfig{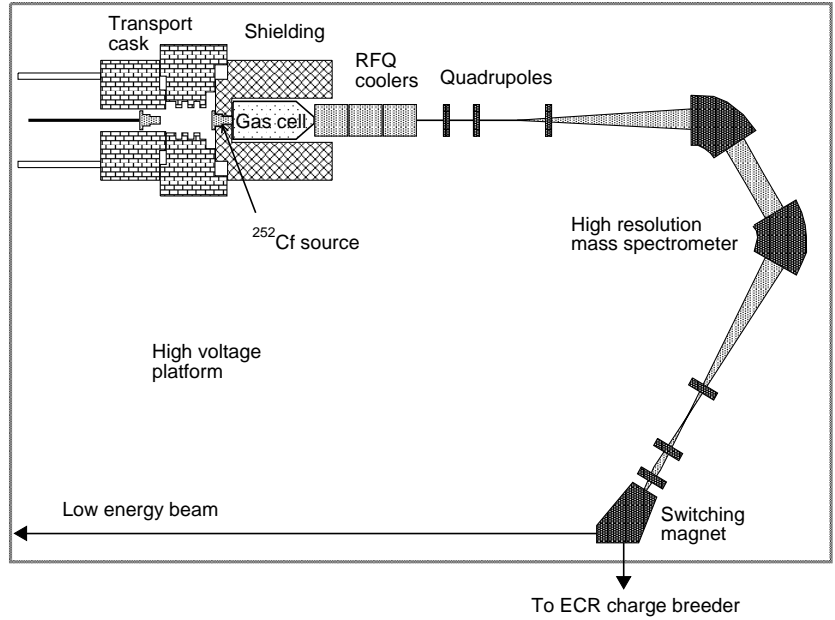}}
\caption{Schematics of the CARIBU radioactive beam injector to ATLAS indicating the location of the cask containing the 1 mCi $^{252}$Cf source, the gas cell, the RFQ cooler, the high resolution mass spectrometer.}
\label{CARIBU}
\end{figure}

The Californium Rare-Ion Beam Utility (CARIBU) at Argonne - presently under construction - will be able to provide precision beams of $^{252}$Cf fission fragments with energies up to 10 MeV/u. This facility employs a 1 mCi $^{252}$Cf source of 5 cm diameter deposited  on a Ta foil facing an RF gas cell~\cite{gascell} of the type used to capture ions at the Canadian Penning Trap (CPT) at ATLAS and proposed as a central component for the future RIA facility. At the exit of the He gas cell, the fragments emerge as 1$^+$ ions, which are subsequently fed into two separated Radio-Frequency Quadrupoles (RFQ). These structures  confine the ions while they are cooled by collisions with the escaping He gas that is pumped away in two stages. After RFQ cooling, the energy spread of the ions is less than $\sim$1 eV and the transverse emittance $\sim$ 3$\pi$ mm mrad. The beam is then  electrostatically accelerated to about 50 keV and the desired nuclei are selected in a high resolution mass separator. The mass separated beam is transported into one of the ECR sources for ATLAS, which in this mode will act as a charge breeder to achieve a high charge state compatible with the ATLAS injector linac PII. All components of the CARIBU source reside on a high voltage platform kept at the same voltage as that for the ECR charge state breeder {\it i.e.} up to $\sim$ 250 kV. The highly charged beams extracted from the ECR breeder will have the same low emittance as stable beams from ATLAS and can be delivered to all experimental stations. A schematic of the CARIBU facility is given in Fig.~\ref{CARIBU}

\section{HELIOS - a novel spectrometer}

\begin{figure}[hbt]
\centerline{\epsfig{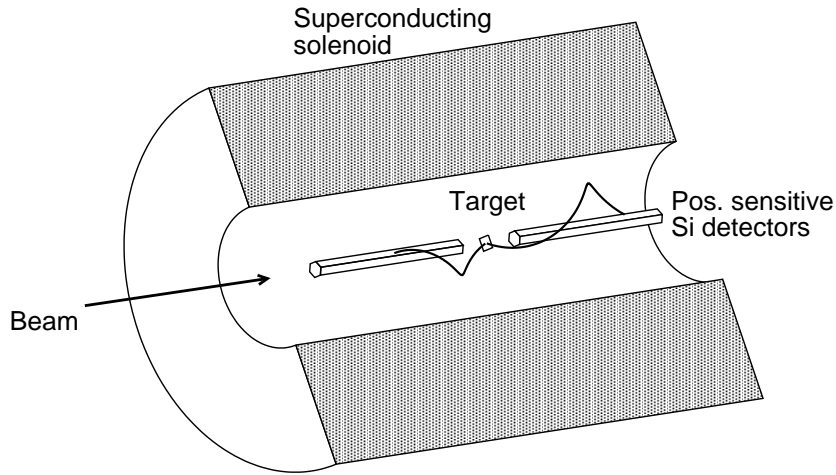}}
\caption{Cut-away schematics of the superconducting solenoid spectrometer. Two helical particle trajectories are indicated by thick solid curves.}
\label{HELIOS}
\end{figure}

In general, radioactive beams can be produced only at significantly lower intensities than those of stable isotopes. Using the in-flight method and the CARIBU injector one can expect  intensities of radioactive beams up to about 10$^6$ part/sec as opposed to the 10$^{10}$-10$^{12}$ part/sec routinely delivered for stable species. This circumstance clearly places a premium on high efficiency detectors such as the HELIcal Orbit Spectrometer, HELIOS, presently being developed at Argonne. A schematic of this device is given in Fig.~\ref{HELIOS}. 
The principle of the spectrometer is to bring the beam in along the axis of the large superconducting solenoid magnet to strike a target near its center~\cite{Schiffer}. Reaction products emitted from the target follow helical trajectories in the uniform longitudinal field inside the solenoid and will either hit the inner wall of the solenoid (if their magnetic rigidity is too large) or return the beam axis. By placing a hollow array of position-sensitive Si detectors around the beam axis it is possible to intercept a large fraction of the particle trajectories and achieve a solid angle for detection that approaches 4$\pi$. From the measured quantities, namely $\Delta T$ - the time-of-flight obtained using the pulsed structure of the ATLAS beam, $z$ - the distance from to target to the intercept with the detector, and $E_{\it lab}$ - the energy deposited in the Si detector, it is possible to derive $m/q$, the mass-to-charge ratio of the particle, $E_{\it cm}$, $\theta_{\it cm}$ its energy and emission angle in the center-of-mass system, as well as a rough measure of the azimuthal emission angle $\phi$. It is an important property of this spectrometer design that the particle energy measured at a specific longitudinal position $z$ exactly reflects the energy dispersion in the center-of-mass system with the consequence that only the intrinsic detector resolution limits the spectral resolution; the kinematical distortions to spectra measured in the laboratory frame-of-reference, that especially for inverse kinematics reactions limit the resolving power of conventional setups, is completely absent in this design.

The main application of this spectrometer is the study of inverse kinematics transfer and inelastic scattering reactions using heavy (radioactive) beams incident on light targets e.g. H,D,$^3$He etc. Such studies promises to obtain information about the structure of neutron-rich nuclei, which cannot be obtained by other means. One favored reaction is H($^{132}$Sn,p), which may be used to determine the characteristics (spectroscopic factors etc.) of single neutron states outside the doubly-closed shell at Z=50, N=82 . 

Some reactions involving heavy ejectiles, e.g. fission fragments, can in principle also be studied in this device although $m/q$ ambiguities may render some applications intractable. One possibility is to use the spectrometer in a gas-filled mode such that fragments, on average, follow trajectories corresponding to the average charge state, $\langle q \rangle$.

\section{Summary}

The radioactive beam program at ATLAS has been reviewed briefly. At present, radioactive beams are predominantly produced via the in-flight method although a sub-set of experiments utilize beams of radioactive species generated at other accelerators. A few representative examples of recent experiments have been presented. These include studies of nuclear reactions of astrophysical interest and studies of the structure of light nuclei. The ATLAS facility is presently being augmented by the inclusion of the CARIBU injector, which will provide beams of fission fragments from a 1 mCi $^{252}$Cf source. The beams obtained in this way overlap only slightly with those from spallation facilities using Uranium targets and will provide a new range of neutron-rich beams for nuclear structure studies. To take full advantage of these new beams, a novel spectrometer (HELIOS) is planned, which will allow for a wide range of inelastic scattering and transfer reaction studies using the neutron-rich beams in inverse kinematics. 
 
\section*{Acknowledgments}
This work was supported by the US Department of Energy, Office of Nuclear Physics, under Contract No. W-31-109-ENG-38.

\vfill\eject
\end{document}